\documentclass[12pt,preprint]{emulateapj}

\usepackage[usenames]{color}

\shortauthors{YAMADA ET AL.}
\shorttitle{Is the black hole in GX 339-4 really spinning rapidly?}

\begin{document} 
\title{Is the black hole in GX 339-4 really spinning rapidly?}

\author{S. Yamada,\altaffilmark{1} K. Makishima,\altaffilmark{1,2}
Y. Uehara,\altaffilmark{1} K. Nakazawa,\altaffilmark{1}
H. Takahashi,\altaffilmark{3}\\
T. Dotani,\altaffilmark{4} Y. Ueda,\altaffilmark{5} K. Ebisawa, \altaffilmark{4}
A. Kubota,\altaffilmark{6}  and  P. Gandhi \altaffilmark{2}\\}

\altaffiltext{1}{Department of Physics, University of Tokyo,
   7-3-1 Hongo, Bunkyo-ku, Tokyo, 113-0033, Japan }
\altaffiltext{2}{Cosmic Radiation Laboratory,
 Institute of Physical and Chemical Research (RIKEN),
   Wako, Saitama, 351-0198, Japan}
\altaffiltext{3}{Department of Physical Science, Hiroshima University,
1-3-1 Kagamiyama, Higashi-hiroshima, 739-8526, Japan}
\altaffiltext{4}{Institute of Space and Astronautical Science, JAXA,
   3-1-1 Yoshinodai, Sagamihara, Kanagawa, 229-8510,  Japan}
\altaffiltext{5}{Department of Physics, Kyoto University,
Kitashirakawa-Oiwake-cho, Sakyo-ku, Kyoto 606-8502, Japan}
\altaffiltext{6}{Department of Electronic Information Systems, Shibaura Institute of Technology,
307 Fukasaku, Minuma-ku, Saitama-shi, Saitama, 337-8570, Japan}

\begin{abstract} 
The wide-band {\it Suzaku} spectra of the black hole binary GX 339$-$4, 
acquired in 2007 February during the Very High state, were reanalyzed.
Effects of event pileup (significant within $\sim 3'$ of the image center)
and telemetry saturation of the XIS data were carefully considered.
The source was detected up to $\sim 300$ keV,
with an unabsorbed 0.5--200 keV luminosity of  
$3.8 \times10^{38}$ erg\,s$^{-1}$ at 8 kpc. 
The spectrum can be approximated by a power-law of photon index 2.7, 
with a mild  soft excess and a hard X-ray hump.
When using the XIS data outside $2'$ of the image center,
the Fe-K line appeared extremely broad,
suggesting a high black hole spin as already reported by Miller et al. (2008) 
based on the same  {\it Suzaku} data and other CCD data.
When the XIS data accumulation is 
further limited to $>3'$ to avoid event pileup,
the Fe-K profile becomes narrower, 
and there appears a marginally better solution that suggests  
the inner disk radius to be $5-14$ times the gravitational radius (1-sigma),
though a maximally spinning black hole is still allowed by the data at the 90\% confidence level.
Consistently, the optically-thick accretion disk is inferred to be 
truncated at a radius $5-32$ times the gravitational radius.
Thus, the {\it Suzaku} data allow an alternative explanation without invoking 
a rapidly spinning black hole.
This inference is further supported by the disk radius
measured  previously in the High/Soft state.
\end{abstract}

\keywords{stars: individual (GX 339-4) --- X-rays: binaries}
\section{INTRODUCTION} 
\label{intro}
One of the current  issues in black hole (BH) research is to measure the spin parameter, $a$. 
This is possible if we can estimate the radius of 
the innermost stable orbit  $R_{\rm in}$,
which decreases from 
$6 R_{\rm g}$ ($R_{\rm g}$ being the gravitational radius) 
down to 1.235  $R_{\rm g}$  as $a$ increases from 0 to 1.
A traditional method of measuring $R_{\rm in}$ of
black-hole binaries  (BHBs) is to parameterize the optically-thick 
disk emission component in their X-ray spectra
in terms of the disk-blackbody ({\tt diskBB}) model (Mitsuda et al. 1984).
First applied successfully to GX~339-4 (Makishima et al. 1986)
and  LMC X-3 (Ebisawa et al. 1993),
this method has been calibrated using  BH masses 
estimated from companion star kinematics, 
and  confirmed to give reliable (e.g., within  $\sim 30$\%)
estimates  of $R_{\rm in}$  (Makishima et al. 2000)
if  distance uncertainties can be neglected.

A more modern way to estimate $a$,
applicable also to active galactic nuclei, 
is to utilize iron line profiles, 
which become broadened and skewed due to stronger relativistic effects 
as $a$ increases (e.g., Fabian et al. 1989).  
First found from the Seyfert galaxy MCG--6-30-15  by {\it ASCA} (Tanaka et al. 1995),
the broad Fe-K lines were later  reported in a number of 
Seyferts using, e.g., {\it XMM-Newton}  (Nandra et al. 2007).
The latest {\it Suzaku} results on MCG--6-30-15
indicate a high spin parameter of $a > 0.917$ (Miniutti et al. 2007). 

Similar high values of $a$ were derived from BHBs (Miller 2007, Miller et al. 2009),
including  GX~339$-$4 observed with {\it XMM-Newton} and {\it Chandra} 
(Miller et al. 2004ab).
Analyzing the {\it Suzaku} data of this BHB acquired in 2007 February,
Miller et al. (2008), hereafter MEA08, 
confirmed the broad Fe-K feature, and argued
that the object is an extreme Kerr BH with $R_{\rm in} \sim R_{\rm g}$.

The wide-band {\it Suzaku} spectra of Cyg X-1 in the Low/Hard state (LHS),
in contrast, gave $R_{\rm in} \sim 15 R_{\rm g} $ (Makishima et al. 2008),
via both the disk emission analysis and the Fe-K line modeling. 
Likewise, $R_{\rm in} \sim 8 R_{\it g}$ was obtained from 
the {\it Suzaku} data of GRO 1655-40 (Takahashi et al. 2008).  
Also, a  2--20 keV {\it Tenma} observation of GX 339$-$4
 (Makishima et al. 1986)  in the High/Soft state (HSS),
with dominant disk emission,
gave $R_{\rm in}= 57\,d_{8}$  km, or  $5.6\,Q$ times  $R_{\rm g}$;
here $Q \equiv d_8/m_7$, $d_8$ is the  distance in 8 kpc (Zdziarski et al. 2004),
$m_7$ is the BH mass in units of a typical value of $7~M_\odot$ (Hynes et al. 2004),
and $R_{\rm in}$ was recalculated applying 
a correction factor of 1.18 (Kubota et al. 1998; Makishima et al. 2000)
and assuming an inclination of $i=30^\circ$ (Gallo et al. 2004).
These results imply  $a \sim 0$.

To examine these apparent discrepancies on $a$,
we reanalyzed the same {\it Suzaku} data of GX 339-4 as MEA03,
and found that the X-ray Imaging Spectrometer (XIS; Koyama et al. 2008) events suffer heavy 
pileup  and telemetry saturation.
These effects, neglected by MEA08,  distort the continuum,
and  {\it indirectly} affect the Fe-K line shape. 

\section{OBSERVATION}
\label{obs}

The present {\it Suzaku} data of GX 339$-$4, as used by MEA08,
were obtained on 2007 February 12,
during an outburst which started in late 2006 (Swank et al. 2006).
The XIS employed the 1/4 window option
and a burst option (0.3 sec for XIS0/XIS1, and 0.5 sec for XIS3), 
to achieve a time resolution of 2 sec and a duty cycle of 15\% or 25\%.
The Hard X-ray Detector (HXD;  Kokubun et al. 2008) was operated in the standard mode.
The data processing and reduction were performed 
in the same way as MEA08,
using the {\it Suzaku} pipeline processing ver.\,2.0.6.13.

In spite of the 1/4 window and burst options,
the XIS events piled up significantly (\S~\ref{instrumental})
as the source was very bright 
($\sim 0.6$ Crab in the  2--10 keV {\it Suzaku} band).
In fact, the unscreened XIS event file contains
an unusually high fraction of Grade 1 events, 
which are produced  when pileup occurs.\footnote{http://www.astro.isas.ac.jp/suzaku/analysis/xis/pileup/ \\ HowToCheckPileup\_v1.pdf.}
In addition, a variable fraction of the CCD frame was
often lost due to telemetry saturation;
this affects the absolute XIS flux.
Out of the 10.3 ks of XIS0 exposure,
only 2.84 ks was free from this problem. 

The HXD-PIN and HXD-GSO spectra, 
acquired for a net exposure of 87.9 ks,
were corrected for small dead times, 
but no other correction due to the source brightness was necessary.
We subtracted modeled non X-ray backgrounds (NXBs; Fukazawa et al. 2009). 
The cosmic X-ray background, $<1\%$ of the total counts, was ignored.
By analyzing the HXD data acquired during Earth occultations,
we confirmed the NXB models to reproduce the PIN and GSO data 
to within 1\%, a typical accuracy level (Fukazawa et al. 2009).
Since the signal becomes comparable to this uncertainty at $\sim$ 300 keV,
we quote the source detection up to $\sim 300 $ keV. 

In the present {\em Letter},
we use data from XIS0, HXD-PIN, and HXD-GSO.
We do not use data from XIS1 or XIS3,
which are affected more by pileup than XIS0.
Unless otherwise stated, errors refer to 90\% confidence limits. 

\section{DATA ANALYSIS AND RESULTS}
\subsection{Effects of event pileup and telemetry saturation}
\label{instrumental}

To examine in detail the XIS0 data for pileup effects, 
we produced a radial count-rate profile
from a 0.5--10.0 keV XIS0 image (discarding telemetry-saturated frames),
and compared it with that of a pileup-free point source,
i.e.,  MCG--6-30-15 observed on 2005 August 17.
The profile ratio between GX 339$-$4 and MCG--6-30-15 
decreases significantly within a radius of $r=2'$ of the image centroid,
reaching at the center $\sim 25\%$ of the ratio at $r>2'$.
Therefore, the XIS0 data are affected by pileup 
 at least within $r< 2'$, and possibly up to $\rm 3'$.

\begin{figure}[bth]
\begin{center}
\vbox{
\includegraphics[scale=0.48]{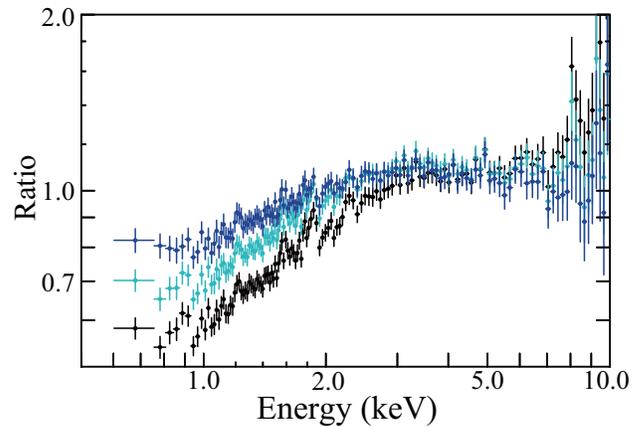}}
\caption{
The XIS0 spectra of GX 339$-$4 taken from different annuli around the source,
$0'-7.5'$ (black), $1'-7.5'$ (cyan), and $2'-7.5'$ (blue), all divided by that over $3'-7.5'$.
The ratios are renormalized to unity at 5.0 keV.
The XIS background, though inclusive, is negligible.}
\label{fig:avespec}
\end{center}
\end{figure}

Figure~\ref{fig:avespec}  shows XIS0 spectra of GX 339$-$4, 
accumulated over different annuli around the source, 
and divided by that outside $3'$ to visualize shape changes.
Pileup thus affects the 1-10 keV continuum shape,
and produces line-like features  at 1.8 and 2.2  keV.
Although MEA08 attributed the line features to response uncertainties, 
it fails to explain the fact that the line strength increases inwards. 
These features, appearing at  the Si-K edge in the XIS 
and the Au M-edge in the X-ray Telescope (Serlemitsos et al. 2007), 
are due to rapid changes of the instrumental response
coupled with pileup. 

\begin{figure}[htbp]
\begin{center}
\includegraphics[scale=0.33]{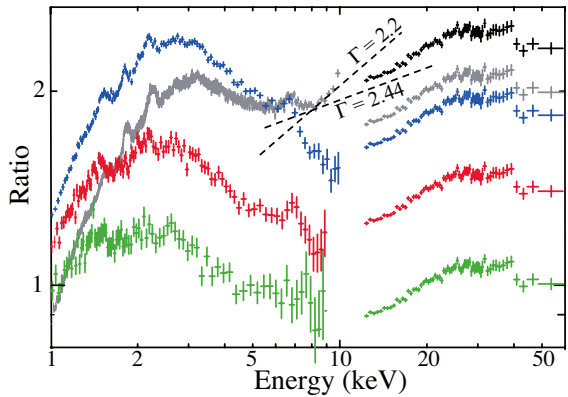}
\caption{ 
The XIS spectra from annuli with different radii, divided by predictions of a common PL model with $\Gamma$ = 2.6;
$0'-7'.5$ (grey), $2'.0-7'.5$ (blue), $3'.0-7'.5$ (red), and $4'.0-7'.5'$ (green).
They are corrected for neither pileup nor telemetry saturation.
The 13--60 keV HXD-PIN spectrum is shown in black.
It is reproduced after scaling by 
live-time fractions of the XIS spectra of the corresponding colors: 
85\%, 79\%, 60\%, and 43\% for grey, blue, red, and green, respectively. 
The two value of $\Gamma$, used in Figure 3, are illustrated. 
}
\label{fig:burst_spectra}
\end{center}
\end{figure}

Figure~\ref{fig:burst_spectra} shows raw XIS0 spectra accumulated 
from different annular regions of the image,
including telemetry-saturated frames,
and corrected for neither pileup nor live-time fraction.
They are shown divided by a powerlaw (PL) prediction,
which is calculated using respective ARFs (Ishisaki et al. 2007) 
that properly consider the fractional photons falling therein. 
In all cases, the PL was chosen to have $\Gamma=2.6$ 
(which approximates the average 2--50 keV spectral slope), 
and a common normalization of 5 ph cm$^{-2}$ keV$^{-1}$ s$^{-1}$, 
absorbed by a column of $N_{\rm H} = 5.0 \times 10^{21}$ cm$^{-2}$. 
Towards the image centroid,
the continua are severely distorted by event pileup.
In addition, the spectral ratio decreases towards outer regions,
which suffer increasingly from the telemetry saturation.
The live-time fractions of XIS0 are 85\%, 79\%, 60\%, and 43\%, 
from the inner to outer annuli. 

In Figure~\ref{fig:burst_spectra}, 
the NXB-subtracted HXD-PIN spectrum is shown in black.  
It is further repeated four times,
but scaled by the XIS live-time fractions
so that a pair of XIS and HXD-PIN spectra with the same color should match.
MEA08 used the grey XIS0 spectrum and the black HXD-PIN data,
but even putting aside the pileup distortions,
this is an incorrect combination.
In MEA08, 
the XIS and HXD-PIN  data points match rather accidentally,
because pileup (which increases the highest XIS spectral end)
and the telemetry saturation (which reduces the XIS normalization)
have opposite effects.

\subsection{Analysis of the $r=0'-4'$ spectra}
\label{corespec}

\begin{figure}[bth]
\begin{center}
\vbox{
\includegraphics[scale=0.2]{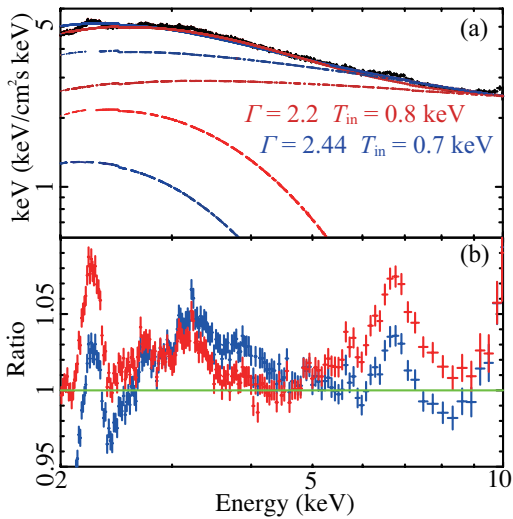}}
\caption{
(a) The XIS0 spectrum of GX 339$-$4 in $\nu F \nu$ form, taken from the $0'-4'$ 
region around the source. 
It is fitted with two different sets of PL+{\tt diskBB} models (see text), ignoring 
the 4-7 keV range. (b) The same spectrum, divided by the two models given in (a). }
\label{fig:pileupspec}
\end{center}
\end{figure}

We tentatively fitted the raw XIS0 spectrum from $0'-4'$
(grey one in Figure~\ref{fig:burst_spectra}),
with a PL plus {\tt diskBB} model,
but ignoring the 4--7 keV  range as MEA08 did.
When using the PL photon index of $\Gamma=2.2$ 
which MEA08 (supplemented by Miller 2009) found for XIS data, 
the {\tt diskBB} temperature was $T_{\rm in}=0.8$ keV (shown in Figure 3 (a) in red).  
The ratio plot (shown in Figure 3 (b) also in red) reveals the broad Fe-K line feature reported by MEA08.
However, the feature becomes much narrower and weaker (blue crosses),
if  we employ $\Gamma=2.44$ 
which is close to the value of $\Gamma=2.4$ used by Miller (2009) for the HXD-PIN data. 
In this case, $T_{\rm in}=0.7$ keV.
Thus, the large Fe-K line width claimed by MEA08 depends on the employed continuum slopes,
which are also indicated in Figure~\ref{fig:burst_spectra}. 
In fact, the slope should not differ by $\Delta \Gamma \sim 0.2$ 
between the XIS and PIN (suzaku memo-2008-06). 

\subsection{Analysis of the $r>2'$ spectra}
\label{bluespec}

Discarding the $r<2'$ region where the pileup effects are severe,
we jointly fit the ``blue" pair of XIS0
and HXD-PIN spectra in Figure~\ref{fig:burst_spectra},
in the 2.4--9.0 keV and 13.0--60.0 keV ranges, respectively.
Though still piled up, 
this XIS0 spectrum ($r=2'-7'.5$) retains high signal statistics.

The spectra in Figure~\ref{fig:burst_spectra} exhibit 
a soft ($\lesssim 8$ keV) and a hard (20--40 keV) hump, 
and an  Fe-K emission line, 
as found before (Zdziarski et al. 1998; Ueda et al. 1993). 
We hence employed a model consisting of a {\tt diskBB}, a PL,
the associated ionized reflection {\tt pexriv} (Magdziarz \& Zdziarski 1994),
and a {\tt laor} (Loar 1991) model for the Fe-K line with the emissivity index fixed at $q=3.0$
and the outer disk radius at $400 R_{\rm g}$. 
The metal abundances in {\tt pexriv}  were fixed 
to solar values except that of iron ($Z_{\rm Fe}$),
while the disk inclination was allowed to vary over $i=25^\circ - 45^\circ$.
The ``constant'' parameter
was set at $C_{\rm XIS}=0.79$ (the live-time fraction), for XIS0 with a tolerance of $\pm 0.05$, 
while at 1.07 (Suzaku memo 2007-11) for HXD-PIN.
We left free all the other model parameters, 
including the absorbing column $N_{\rm H}$.

\begin{figure}
\begin{center}
\vbox{
\includegraphics[scale=0.42]{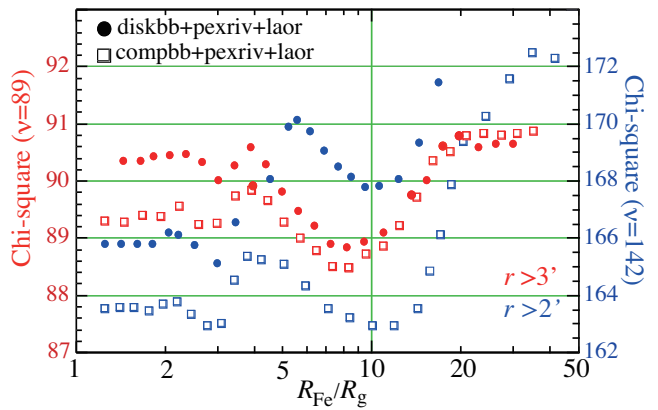}}
\caption{Chi-squares of the joint fits to the XIS0 and HXD-PIN spectra,
shown as a function of the {\tt laor} inner radius $R_{\rm Fe}$.
Blue and red data points use the spectra 
of the same color as in Figure~\ref{fig:burst_spectra}.
Filled circles show the absorbed {\tt diskbb+pexriv+laor} fit,
with the degree of freedom $\nu$ given on the ordinate.
Open squares are derived when {\tt diskbb} is replaced with {\tt compbb}, 
with $\nu$ decreasing by 2. 
}
\label{fig:chisq}
\end{center}
\end{figure}

The fit was acceptable,
and behaved as shown by filled blue circles in Figure~\ref{fig:chisq}
as a function of the  {\tt laor} inner radius, $R_{\rm Fe}$.
We thus find $R_{\rm Fe}<3.6 R_{\rm g}$,
which is consistent with the conclusion of MEA08.
The  other parameters are: 
$C_{\rm XIS} = 0.78 \pm 0.02$,
$\Gamma=2.68 \pm 0.02$, $T_{\rm in}=0.73^{+0.03}_{-0.02}$ keV,
the {\tt pexriv} solid angle fraction $\Omega/2\pi =0.64^{+0.02}_{-0.03}$,
its ionization parameter $\log \xi = 2.53^{+0.38}_{-0.74}$,
and  $Z_{\rm Fe} = 1.77 ^{+0.13}_{-0.15}$.
Fixing $Z_{\rm Fe} = 1.0$ worsened the fit by $\Delta \chi^2 \sim 30$.
The {\tt laor} rest-frame energy was  
$E_{\rm c}=6.71^{+0.09}_{-0.53}$ keV, and its  inclination $i = 43^\circ~^{(+2)}_{-4}$.
However, the  Fe-K equivalent width (EW),
$230^{+80}_{-40}$ eV,  much exceeds 
the value of $\sim 110$ eV (George \& Fabian 1991)
predicted by the derived $\Gamma$, $\Omega/2\pi$, and $Z_{\rm Fe}$.

The steep slope of $\Gamma \sim 2.7$ indicates
that  the source was in the Very High state (VHS; Miyamoto et al. 1993),
wherein the disk emission must be Comptonized (Kubota \& Makishima 2004).
We hence replaced {\tt diskbb} with a Comptonized blackbody, {\tt compbb};
the photon source it assumes, a single  blackbody,
may be reconciled with the multi-color  {\tt diskbb} formalism by the free $N_{\rm H}$.
The results of this analysis are given in Figure~\ref{fig:chisq}
by open blue squares.
Thus, the local $\chi^2$ minimum at  $R_{\rm Fe} \sim 10 R_{\rm g}$,
which existed in the  {\tt diskbb} modeling,
became as good as the small-$R_{\rm Fe}$ solution.
This large-$R_{\rm Fe}$ solution is characterized by 
a blackbody temperature of $0.54 \pm 0.01$ keV,
a Compton optical depth of $\tau = 1.09^{+0.06}_{-0.12}$,
an electron temperature of $9.5^{+0.9}_{-1.6}$ keV,
$E_{\rm c} = 6.29^{+0.22}_{-0.18}$ keV and  $i = 42^\circ~^{+3}_{-11}$.
Again, we find that the Fe-line solutions are degenerate,
depending on the continuum choice.

\subsection{Analysis of the $r>3'$ spectra}
\label{redspec} 

We finally analyze the ``red"  spectra in Figure~\ref{fig:burst_spectra}
in the same way as in \S~\ref{bluespec},
under a constraint of $C_{\rm XIS}= 0.60\pm 0.06$.
Although this XIS0 spectrum 
($r=3'-7'.5)$ is estimated to be still weakly ($\lesssim 10 \%$) piled up,
we do not  correct for this,
because no established  method  is available yet.
The HXD data are the same as in \S~\ref{bluespec}.
In advance, the red XIS0 spectrum and that of XIS3 from $r=3.5'-7.5'$,
where the count rates are comparable,
were confirmed to have a constant ratio within $\sim 10\%$.
We also confirmed with the MCG--6-30-15 data
that XIS spectra from these outer regions 
have a constant (within $\sim 10\%$) ratios to those from the image core.
 
As shown in Figure~\ref{fig:chisq} in red,
the results from the {\tt diskbb} and {\tt compbb} modeling
agree better than for $r >$ 2, 
with $\tau = 0.16^{+0.55}_{-0.16}$ for  {\tt compbb}.
Below, we examine  the {\tt diskbb}+{\tt pexriv}+{\tt laor} fit
(filled red circles). 
In contrast to the ``blue" spectra (\S~\ref{bluespec}),
the data now prefer the small-$R_{\rm Fe}$ solution as
$5.0< R_{\rm Fe}/R_{\rm g} <14$
(with minimum at 8.2) at  68\% confidence,
although $R_{\rm Fe}$ is unconstrained at  90\% 
level due to poor statistics.
The other parameters  are: 
$T_{\rm in}=0.72 ^{+0.04}_{-0.03}$ keV,
$\Gamma=2.67 ^{+0.06}_{-0.02}$,
$\Omega/2\pi =0.60 \pm 0.02$,
$Z_{\rm Fe} = 1.71 ^{+0.17}_{-0.13}$,
$N_{\rm H} = 0.55^{+0.27}_{-0.13}\times 10^{22} $ cm$^{-2}$,
and $i \sim 33 ^\circ$ (the entire range of $25^\circ$--$45^\circ$ is allowed at the 90 \% confidence).
 The line energy $E_{\rm c}=6.68 ^{+0.41}_{-0.47}$ keV
is consistent with the derived range of 
$\log \xi = 3.65^{+\infty}_{-1.78}$.
The Fe-line EW, $104^{+67}_{-53}$ eV, is 
also now in agreement with the prediction ($\sim 100$ eV) 
from $\Gamma$, $\Omega/2\pi$, and $Z_{\rm Fe}$.
Employing $i=30^\circ$ and  the correction factor 1.18 (\S~\ref{intro}),
the {\tt diskbb} normalization yields
$R_{\rm{in}}  = (57^{+11}_{-15}) \;d_{8}$  km,
or  $R_{\rm in}/R_{\rm g}= (5.6^{+1.0}_{-1.5})\; Q$.
The results remain largely unchanged when fixing $Z_{\rm Fe}$ at 1.0.

\begin{figure}
\begin{center}
\vbox{
\includegraphics[scale=0.34]{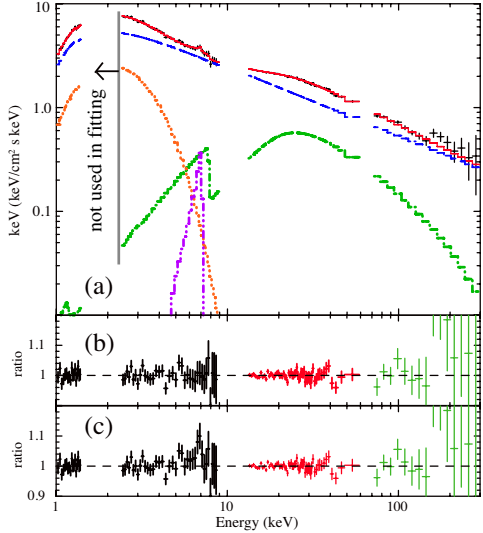}}
\caption{
(a) The XIS0 ($r=3'-7'.5$), HXD-PIN, and HXD-GSO spectra (black crosses),
simultaneously fitted with the absorbed {\tt diskbb+pexriv+laor} model (red)
and presented in a deconvolved $\nu F \nu$ form. 
The PL (blue), reflection (green), 
{\tt diskbb} (orange), and   {\tt laor} (purple) components are also shown. 
(b) The fit residuals.
(c) The same as panel (b), but the {\tt laor} component is set to 0.
}
\label{fig:specfit}
\end{center}
\end{figure}

As shown in Figure~\ref{fig:specfit}(a), the best-fit model has been 
extended  to incorporate the 70--300 keV HXD-GSO data,
resulting in an acceptable ($\chi^2/\nu = 108.7/102$) simultaneous fit.
The model parameters have remained unchanged within their statistical errors.
Panels (b) and (c)  therein show the fit residuals.
We tried a simple continuum blurring, and found no important change, but we defer detailed analysis to 
later work.

\section{DISCUSSION}
\label{discussion}
We reanalyzed the  {\it Suzaku}  XIS0 and HXD data of GX 339$-4$ 
acquired in the 2007 VHS.
When the disk emission was modeled with {\tt diskbb},
the ``blue" XIS0 spectrum, using $r>2'$,
gave a small  Fe-K  inner radius as $R_{\rm Fe}< 3.6 R_{\rm g}$ (\S~\ref{bluespec}).
Although this  apparently reconfirms the high BH spin by MEA08,
the Fe line EW is too large (\S 3.3).
In addition, a large-$R_{\rm Fe}$ ($\sim 10 R_{\rm g}$) solution is also allowed
when considering disk Comptonization that is usually the case in the VHS.
Thus, the Fe line shapes degenerate,
depending on the continuum modeling.

To further eliminate the XIS pileup effects,
we utilized the ``red" XIS0 spectrum from $r>3'$  (\S 3.4).
Then, the {\tt diskbb} and {\tt compbb} results became more consistent.
Importantly, the large-$R_{\rm Fe}$ solution (e.g., $5.0< R_{\rm Fe}/R_{\rm g} <14$ at $1\sigma$), 
are preferred, although the small-$R_{\rm Fe}$ solution still remains valid at the 90\% 
confidence level.
%
The implied Fe-K line EW and the center energy 
are consistent with the reflection parameters (\S 3.4). 

Assuming isotropic emission,
the 0.5--200 keV luminosity is 
$3.8 \times 10^{38}\,d_8^2$ erg s$^{-1}$,
or $\sim 0.4\,d_8^2m_7^{-1}$ times the Eddington limit ($L_{\rm Ed}$).
Since the VHS have so far been observed in other sources over a typical
luminosity range of $ (0.2-1) L_{\rm Ed}$ (Kubota \& Makishima 2004),
this indicates $ 0.5 < d_8^2/m_7 < 2.5$,
or $ 0.7/\sqrt{m_7} < Q < 1.3/\sqrt{m_7}$.
Considering extreme BH masses of 
$3\;M_\odot$ ($m_7= 0.43$) and  $15\;M_\odot$ ($m_7= 2.1$),
we then obtain  $0.5 < Q < 2$. 

The  {\tt diskbb} radius $R_{\rm in}/R_{\rm g}=(5.6^{+1.0}_{-1.5}) \; Q$, 
found in \S~3.4, is in fact a lower limit,
because a significant fraction of disk photons will be  Comptonized 
into the PL by hot electron clouds (Kubota \& Makishima 2004).
Then, the true disk area must be a sum of the {\tt diskBB} area
and that of  the seed photon source.
Since the 0.5--200 keV photon number in the PL component is 
$\sim5$ times larger than that contained in the 0.5--10 keV {\tt diskBB} emission,
the estimated radius will increase by a factor of  $\sqrt{1+5}$,
to $R_{\rm in}/R_{\rm g} =(10-16)\; Q$.
The above estimated uncertainty in $Q$
then yields $R_{\rm in}/R_{\rm g} =(5-32)$. 
Since we derived this range assuming $i = 30^{\circ}$ (or $\sqrt{\cos i }$ = 0.93), 
uncertainties in $i$ can increase the upper bound, but would not affect the lower bound by 
more than $\sim$ 7\%.

The present  Fe-line and {\tt diskbb} analyses consistently suggest 
$R_{\rm in}/R_{\rm g} \gtrsim 5$,
and hence $a<0.4$, in contrast to the value of $a = 0.89 \pm 0.04$ by MEA08.
If our interpretation is correct, GX 339-4 is inferred to be spinning only mildly (if at all),
like Cyg X-1 (\S~\ref{intro}; Miller et al. 2009).
In addition, the consistency between the two methods reconfirms
that $Q$ is relatively close to unity.

We must admit that the assumption of 
a single PL continuum down to $\sim 1$ keV might not be warranted.
In that sense, the value of $R_{\rm in}/R_{\rm g}= 5.6\,Q$ (\S~\ref{intro})
measured with {\it Tenma} is more reliable,
since it was obtained  in the  HSS 
where the spectral modeling is much less ambiguous,
and by a non-CCD instrument that is free from any pileup.
This, together with our estimates on $Q$,
further argues against the large spin parameters.

So far, large values of $a$ were reported for GX~339$-$4 
based on {\it Chandra} data,
and from  {\it XMM-Newton} plus {\it RXTE}  (Miller et al. 2004ab, and Reis et al. 2008).
However, these measurements have not been examined for
systematic effects due to the continuum choice employed therein.
In addition, the former may be limited by the continuum bandwidth,
and the latter could be still subject to CCD pileup.

\smallskip
The authors would like to express their thanks to the Suzaku team members and an anonymous referee who gave us valuable comments.
The present work was supported by Grant-in-Aid for JSPS Fellows.











\end{document}